# A Systematic Review of Digital Transformation Literature (2013 – 2021) and the development of an overarching a-priori model to guide future research

## Full research paper

**Mekhala Egodawele**
Faculty of Business, Law and Arts
Southern Cross University
Queensland, Australia
Email: m.adhikari.egodawele.11@student.scu.edu.au

**Darshana Sedera**
Faculty of Business, Law and Arts
Southern Cross University
Queensland, Australia
Email: darshana.sedera@scu.edu.au

**Vinh Bui**
Faculty of Science and Engineering
Southern Cross University
Queensland, Australia
Email: vinh.bui@scu.edu.au

## Abstract

In recent times, organizations purport to undergo unprecedented transformations owing to the adoption of digital technologies. Consequently, there has been a substantial effort in academia attempting to better understand the phenomenon of digital transformation in business organizations. However, a cumulative tradition of research on digital transformation, underpinned by a consolidated theoretical positioning, is compromised by the loosely defined constructs, confusion in terminology and lack of an overarching framework of its nomological net. This paper, therefore, features a systematic review of the assorted and fragmented literature on this notion of Digital Transformation by critically analysing 174 peer-reviewed journal articles published between 2013 and 2021, in over thirty leading academic outlets. The authors provide a consolidated nomological net of digital transformation by synthesizing themes and dominant theories apparent in existing digital transformation literature, which will be useful for future academic studies.

# 1　Introduction

Digital transformation is "an evolutionary process that leverages digital capabilities and technologies to enable business models, operational processes and customer experiences to create value" (Morakanyane et al. 2017, p. 9). Digital technologies have overhauled how organizations conduct business and offered myriad opportunities to deliver digital innovations (Fichman et al. 2014; Sedera and Lokuge 2017). Leong et al. (2017, p. 97) assert that "technology presents itself as the tool and the means for the firm to not only survive but thrive to date." The adoption of ubiquitous digital technologies such as mobile devices and the Internet of Things (IoT) together with the power of social media, cloud computing, big data analytics and artificial intelligence prompts the digital transformation of organizations, reinforcing their products, services, business processes and business models (Lokuge and Sedera 2016; Lokuge et al. 2019; Sebastian et al. 2017; Tan et al. 2016; Vial 2019). Digital technologies possess salient and distinct characteristics such as flexibility, transferability, edibility, decomposability, traceability, interoperability, availability, scalability, reliability, security, adaptability, and reduced cost that deliver a wealth of favourable opportunities to organizations (Nylén and Holmstrom 2017; Sedera et al. 2016; Tan et al. 2016; Wessel et al. 2020). In fact, these disruptive technologies act as operand and operant resources (Nambisan 2013). As such, digital transformation entails "the creation of, and the consequent change in, market offerings, business processes, or models that result from the use of digital technology" (Nambisan et al. 2017, p. 224) and these changes are required to be well tackled. The triumph of organizations such as Amazon and Netflix and the failure of conventional organizations such as Blockbuster can be cited as examples of digital transformation (Goh et al. 2011). Wessel et al. (2020, p. 103) emphasize that in digital transformation, "digital technology is central to redefining value propositions, which leads to the emergence of a new organizational identity," while Tan et al. (2020) posit how technologies facilitate effective business processes through the digital transformation of the business ecosystem.

Albeit the extensive digital transformation literature, there is a dearth of a consolidated view on how digital transformation occurs and attention to its broader nomological net (Warner and Wäger 2019). As such, there exists an opportunity in the literature to better understand the notions of digital transformation as an input-process-output model, explained by appropriate theories, using its complete nomological framework. A similar work was completed by Gastaldi et al. (2018) where they portrayed digital transformation as a process of antecedents, processes and outcomes, albeit limited to the health care sector. Lokuge and Duan (2021) also illustrate the common use of terminologies such as digitization, digitalization and digital transformation in academia and industry. Moreover, the terms of digitization, digitalization and digital transformation have created a degree of confusion, that require clarity (Verhoef et al. 2021). Similarly, the research community will benefit from a consolidated theoretical exposition of digital transformation. This paper aims to provide a state-of-the-art literature synthesis on digital transformation and derive an inclusive nomological net that illustrates a consolidated view that theoretically underpins this multifaceted notion. In doing so, the study also clarifies the theoretical positioning of the resources, their configuration, outcomes, capabilities, context and considerations under the umbrella of the three phases of digital transformation. As such, this research paper attempts to derive a nomological net of digital transformation research, explained by the salient existing theoretical views. Consistent with the research aim, the authors systematically review 174 peer-reviewed articles on digital transformation gathered from Information Systems and Management Science research published between 2013 – 2021. The authors believe that this study invites prospective researchers to view digital transformation as an evolving phenomenon while adding value to the existing literature. Despite the ubiquity instilled in digital transformation and the myriad of studies reviewing past digital transformation literature, academic literature has so far paid surprisingly less attention to the development of a consolidated view on the notion of digital transformation underpinned by prominent theoretical perspectives in the existing literature.

The paper proceeds in the following manner. The following section discusses the methodology of the literature synthesis. Next, the results are analysed, making several vital observations. The conclusion section summarizes the findings of the study. Finally, the paper derives an a-priori nomological net model, and the concluding section discusses limitations, practical implications and avenues for future research in relation to the study.

# 2　Methodology

A good literature review will generate a sound foundation for understanding the former discoveries made by the previous scholarly work in the same discipline (Saunders et al. 2019). Initially, the authors identified the research objectives and (refer to section 1) and developed the literature review protocol that featured the criteria used in the research document selection. As cited by Tranfield et al. (2003), the protocol confirmed the research keywords and terms according to the scope of the review and the past discussions. The articles in non-English languages and published before January 2013 and after December 2021 and those published in conference proceedings, book reviews, dissertations, case studies and books were deemed under the exclusion criteria in the protocol. The authors also decided to select papers published in reputable journals for analysis using the Australian Business Deans Council (ABDC) rankings list and the Scimago classification as a guideline. The sources were limited to the top-tier peer-reviewed scholarly journal articles in leading academic outlets owing to their high impact in research fields (Podsakoff et al. 2005) and papers of high quality, while the criterion has been used in other literature review studies in the past as well (Crossan and Apaydin 2010). The year 2013 was selected as the baseline for this study as it was the year of digital business that saw the emergence of the notion of digital transformation with the introduction of digital technologies such as social, cloud, big data and mobile (Fenwick 2012). English articles were selected due to the abundance of publications written in English that allowed the authors to get a wide range of sources for the review.

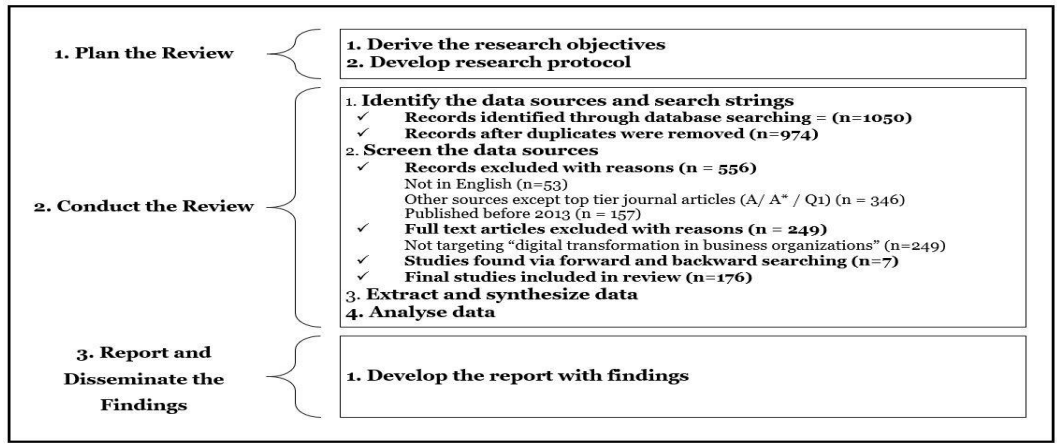

*Figure 1: Phases of the systematic literature review (adapted from Tranfield et al. 2003)*

The authors used the search string with Boolean operators between keywords and search fields. The phrases: Title: ("digital transformation" OR "digitalization" OR "digital*" OR "digitiz*" OR "digitization" OR Abstract: ("digital transformation" OR "digitalization" OR "digital*" OR "digitiz*" OR "digitization" OR Keywords: ("digital transformation" OR "digitalization" OR "digital*" OR "digitiz*" OR "digitization"). The asterix enabled the authors to identify a wide range of sources with titles, abstracts and keywords with various endings. The authors conducted a search using the above strings in the abstract, title and keywords of the articles indexed in databases like EBSCO Business Source Complete, SAGE, Scopus, Wiley Online Library, Emerald journals, Web of Science and Springer Link as they provide users with access to a broad number of top tier peer-reviewed IS journals and others set in a plethora of research fields. The authors screened the titles and abstracts of the journal articles against the inclusion criteria to decide whether the article requires further reading or should be excluded (Okoli et al. 2010) (see Figure 1). The screening of articles eventuated in a total of 174 articles. This sample was deployed to expedite the other phases of the literature review, and this was done after examining the full papers to assure their conformity with their inclusion criteria (Liberati et al. 2009). Due to the limitation of space, the following table (Table 2) summarizes the journals that consisted of more than 5 papers while those with less than 5 papers are classified as 'other.'

| Journal | 2013 | 2014 | 2015 | 2016 | 2017 | 2018 | 2019 | 2020 | 2021 | Total |
|---|---|---|---|---|---|---|---|---|---|---|
| BPMJ | 0 | 0 | 0 | 0 | 2 | 3 | 3 | 0 | 1 | 9 |
| CMR | 0 | 0 | 0 | 0 | 1 | 0 | 1 | 7 | 2 | 11 |
| IJEBR | 0 | 0 | 0 | 0 | 0 | 1 | 0 | 1 | 4 | 6 |
| IJIM | 0 | 0 | 0 | 0 | 1 | 0 | 0 | 5 | 3 | 9 |
| IJOPM | 0 | 0 | 0 | 0 | 0 | 0 | 0 | 2 | 4 | 6 |
| IMM | 0 | 0 | 0 | 0 | 1 | 0 | 0 | 8 | 2 | 11 |
| JBR | 0 | 0 | 0 | 0 | 0 | 0 | 5 | 3 | 11 | 19 |
| JEIM | 0 | 0 | 0 | 0 | 0 | 0 | 1 | 1 | 3 | 5 |
| JSIS | 0 | 0 | 0 | 0 | 0 | 1 | 2 | 4 | 2 | 9 |
| MISQ | 8 | 0 | 0 | 1 | 0 | 0 | 0 | 0 | 1 | 10 |
| MS | 2 | 1 | 2 | 4 | 5 | 4 | 2 | 1 | 2 | 23 |
| RTM | 0 | 0 | 2 | 1 | 0 | 2 | 1 | 2 | 1 | 9 |
| TFSC | 0 | 0 | 0 | 0 | 0 | 0 | 2 | 1 | 8 | 11 |
| Other | 0 | 1 | 1 | 3 | 4 | 4 | 6 | 11 | 6 | 36 |
| **Total** | **10** | **2** | **5** | **9** | **14** | **15** | **23** | **46** | **50** | **174** |

*Table 1. Summary of Articles*

The authors categorized the articles into pre-determined themes based on their understanding of the current theories and the variables (Levy and Ellis 2006). The well-renowned structuring content analysis approach of Mayring (2014; 2004) facilitates an organized and theory-guided reduction of a wide range of data from any source to gist by categorizing the data into common themes (categories) (Fastenrath and Braun 2018). The authors selected this approach to complete the analysis and they followed five steps namely (i) develop a category system according to the research questions, (ii) code relevant passages in the text as per the category system, and (iii) revise previously developed classification framework, (iv) code the text according to the revised category system and (v) interpret and discuss the results. Hence, the authors deductively generated the major themes and refined the categories to create sub-themes, following a pilot review they conducted considering about 20% (35 articles) (Mayring 2014). The category system was further refined when the analysis covered about two third of the sample (115 papers), where some subcategories were further divided (Mayring 2014, p. 95).

## 3 Findings

### 3.1 Analysis of Publication Trends

The authors identified the trends of digital transformation research by discussing the distribution of the reviewed articles by year. It was apparent that the highest number of peer-reviewed digital transformation articles was published in 2021. There had been a dramatic rise in the publications from 2015, following a slump in 2014. The years 2020 and 2021 have seen the highest percentage of publications based on digital transformation. These trends portray the burgeoning recognition of the phenomena and digital business transformation worldwide as the new decade dawned, as acknowledged by renowned organizations such as (BCG 2022; McKinsey 2020). Perhaps, this could be a consequence of the pandemic and post-pandemic era notably featured during the period (Stackpole 2021). The rate of publications from 2014 to had been substantially slower than that portrayed after 2019.

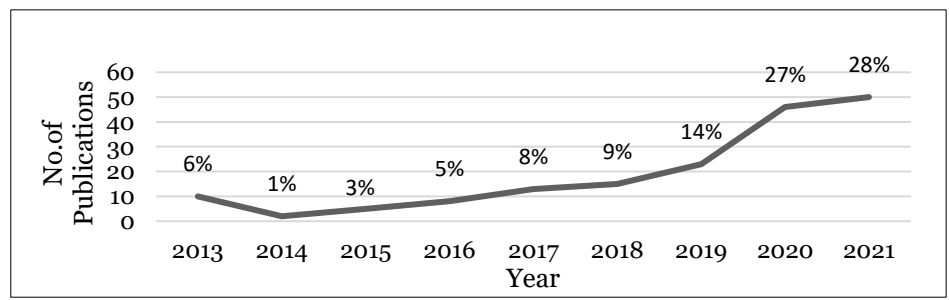

*Figure 2: Articles on Digital Transformation published over time (2013 – 2021)*

### 3.2 Dominant Theoretical Foundations Evidenced in the Literature

The authors identified four dominant theoretical foundations which surfaced throughout the review to build a consolidated theoretical framework on the digital transformation in business organizations. It was apparent that several studies have adopted theories such as the resource-based view, configuration theory, dynamic capabilities theory and ambidexterity theory to address many notions related to digital transformation phenomena.

#### 3.2.1 Resource-Based View

The resource-based view of the firms can be traced back to the year 1959 where Penrose (1995), showcased that firms consist of a set of resources while management research aids and limits the growth of the firms for the ideal deployment of the current resources. The RBV has been extensively used in digital transformation literature to elucidate how organizations survive with a sustainable competitive edge (Barney 1991). RBV states that a firm gains a competitive advantage utilizing the fusion of valuable, rare, imperfectly imitable and non-substitutable (VRIN) resources and capabilities (Barney 2001; Bharadwaj 2000). As per Verhoef et al. (2021), the digital transformation being a multidisciplinary notion that encompasses the facets of the organization, supply chains, information technology, strategy and marketing, RBV could be an ideal theoretical lens via which digitalization, particularly concerning the small and medium enterprises (SMEs) where the three primary SME resources are cited as digital strategy, information technology and employee skills. Chen et al. (2016) portray the significance of IT capability through the lens of RBV, where it is deemed to possess attributes that can enhance the performance of a firm. Pergelova et al. (2018) combine the RBV of the firm with a view based on cognition to assess the impact of digital technologies that are mediated by global market intelligence on the internationalization of SMEs.

### 3.2.2 Dynamic Capabilities

As per Helfat and Raubitschek (2018), this theory posits how firms create and sustain competitive advantage. As gestated by Teece (2014), the dynamic capabilities possess ordinary capabilities such as operational, administrative and governance-related capabilities that are required to fulfil tasks and dynamic capabilities, which include those tasks that allow a firm to direct the ordinary tasks to the efforts that generate more incredible value. Mazumder and Garg (2021) portray the dynamic capabilities of service providers, namely consultative, orchestration, insights, network management, knowledge access, and standardization that facilitate digital transformation. This theory is an extension of the RBV of the firm where it is based on the firm's ability to change its resource base to ensure its survival while strengthening the extent of fitness with the firm environment (Jiang et al. 2015). The importance of dynamic capabilities in facilitating strategic change and digital servitization amidst a competitive, digital and customer0driven business context has been emphasized by (Coreynen et al. 2020). Soluk and Kammerlander (2021) demonstrate that the three phases they identified, namely, process digitalization, product and service digitalization, and business model digitalization, require dynamic capabilities throughout.

### 3.2.3 Configuration Theory

Strategy is asserted as the scope and direction of a firm in the long run that reaps advantages in a dynamic environment via its configuration of competencies and resources to achieve the expectations of its stakeholders (Scholes et al. 2002). The configuration theory deems that the whole is ideally understood from a system view and needs to be perceived as an interconnected pattern of elements. Henry Mintzberg supported a configurational approach to the theory of management and organization. This theory portrays configurations as constellations of mutually supportive features such as organizational structures, processes and strategies (Miller and Friesen 1978). As per Reck and Fliaster (2019), configurations result in accelerated digitization of business organizations where the configurations portray three patterns of interaction among the skills, networks and behaviours of the chief digital officers. Lanzolla et al. (2020) posit that configurational theory will likely aid in comprehending the fusions where the digital transformation benefits are either formal design mechanisms or a substitute or complementary.

### 3.2.4 Ambidexterity Theory

Preceding literature on digital transformation portrays the impact of digital transformation, particularly in relation to the firm's performance, where several scholars also feature organizational ambidexterity in their studies (Li et al. 2018). Exploitation involves existing internal resources and exploration involves discovering novel capabilities and knowledge (Cenamor et al. 2019). Organizational ambidexterity occurs when a firm accomplishes exploration and exploitation simultaneously (Adler et al. 1999). Scholars have addressed the deployment of digital technologies in achieving ambidexterity via exploiting the current resources while exploring digital innovation (Raisch et al. 2009). Chang and Hughes (2012) portray that ambidexterity is more appropriate for firms that confront excessive resource constraints) and it is apparent that balanced exploration and exploitation of resources allow firms facing resource scarcity to enhance their performance (Cao et al. 2009). Considering ambidexterity and the execution of a digital transformation, the need to act promptly and explore while managing a conventional enterprise is required. The use of digital technologies in a firm ensures the balance between the conventional competing goals, which is a crucial capability that resonates with ambidexterity in a firm (Gibson and Birkinshaw 2004).

## 3.3 Theoretically Driven Nomological Net of Digital Transformation

Using the theoretical foundations that surfaced in the above section, the authors identified the process of digital transformation in business organizations in terms of resources (R), capabilities, considerations and contextual factors (C3) that are being configured and reconfigured to reap a range of outcomes (O). For instance, it was apparent that a unique configuration of resources and capabilities fused with objectives to reap superior performance eventuates in a firm's competitive advantages (Bacon et al. 2019; Park et al. 2012). A nomological net is an interlocking network of laws that add to a theory (Cronbach and Meehl 1955) and the authors propose the RC3O nomological net of digital transformation (see figure 3) following the identification of the thematic strands and the subthemes portrayed in Table 2.

| Themes | Sub-themes |
|---|---|
| Resources impacting on digital transformation 100 | Resource Utilization (Skylar et al. 2019; Li et al. 2020) |
| | Resource Availability (Shashi et al. 2020; Nelson et al. 2017; Becker and Schmid 2020; Euchner 2019) |
| | Digital Expertise (Rech and Fliaster 2019; Maedche 2016; Matt et al. 2015; Ardito et al. 2018; Nelson et al. 2017) |
| | Digital Leadership (Legner et al. 2017; Weil et al. 2021) |
| Capabilities impacting how the resources are configured for digital transformation | Digital Responsibilities (Li 2020; Rech and Fliaster 2019; Alalwan et al. 2021; Bharadwaj et al. 2013; Westerman 2016) |
| | Strategic Expertise (Rech and Fliaster 2019; Correani et al. 2020) |
| Considerations impacting the configuration of resources for digital transformation | Economic (Soluk and Kammerlander 2021; Drechsler et al.2020) |
| | Social (Papagiannidis et al.2020; Li 2020) |
| | Technological (Ferreira et al. 2019; Fritzgerald et al. 2013) |
| | Environmental (Lokuge et al. 2021; Ross et al. 2017) |
| | Internal (Chanias et al. 2019; Björkdahl 2020; Kraus et al. 2018; Lanzolla et al. 2020; Fletcher and Griffiths 2020) |
| Contextual factors impacting resource of configuration for digital transformation | Relative (Björkdahl 2020; Solberg et al. 2020; Wang et al. 2020; Fletcher and Griffiths 2020) |
| | Definitive (Woodard et al. 2013; Ukko et al. 2019; Michelman 2018; Ribeiro-Navarrete et al. 2021) |
| Outcomes of resource configuration for digital transformation | Positive Outcomes (Paschou et al. 2020; Gavrila and Ancillo 2021; Michelman 2018; Euchner 2019; Naik et al. 2020; Sklyar et al. 2019; Lokuge et al. 2021) |
| | Negative Outcomes (Lanzolla et al. 2020; Cennamo et al. 2020; Schaarschmidt et al. 2021; Galindo-Martín et al. 2019; Legner et al.2017; Laurenza et al.,2017) |

*Table 2. Themes and Subthemes*

### 3.3.1 Resources impacting on digital transformation

*"Valuable firm resources possessed by large numbers of competing or potentially competing firms cannot be sources of either a competitive advantage or a sustained competitive advantage." (Barney 2001, p. 106).*

The analysis revealed 2 sub-themes of resources and as listed in Table 2, they include resource utilization and resource availability. Optimized allocation of resources per the prioritization of company goals is crucial for digital transformation (Griva et al. 2021). Chen et al. (2021) relate to the resource sharing approach among partners to steer clear of incompatibilities and it is apparent that a comprehensive digital strategy could be expedited when digital resources are used across the firm without restricting the resources to the function of IT (Bharadwaj et al. 2013). A number of studies surveyed in the review identified the availability and the role of technological, intellectual, human and financial resources (Becker and Schmid 2020; Nelson et al. 2017; Sedera 2006) in the organizations. Kindermann et al. (2021) deploy the RBV theory to assess the kinds of digital resources that should be combined to achieve sustainable competitive advantage in a firm. Becker and Schmid (2020) gestate the availability of sufficient financial resources as an antecedent of digital transformation while Olsson et al. (2020) demonstrate the lack of resources as a factor that imposes a detrimental effect on digital transformation. Their findings recognize the presence of a resource constraint specifically among female entrepreneurs. Canhoto et al. (2021) portray the same observation in terms of financial and human resources among the SMEs and suggests reconfiguration as an appropriate approach to align them digitally. Keller et al. (2021) showcase the high cost incurred by the digitalization of the business processes and past literature portray several aspects that resonate with the lack of digital resources such as the digital divide that exists among the SMEs and large-scale organizations and lack of access to resources (Pergelova et al. 2018).

### 3.3.2 Effect of capabilities on resource configuration for digital transformation

*"The next ten years will bring fundamental changes to our working world, and to adapt, employees in almost every role and industry will need to acquire new skills." (McKinsey 2021, p. 1).*

This review revealed six subthemes that include digital expertise, digital responsibilities, digital capabilities, interdisciplinary collaborations, strategic expertise and ambidextrous attitude. In their study, Park and Mithas (2020) portray the significance of the configuration of capabilities to generate an enhanced idea of the role of information technology in digital business transformation. Digital fluency, technological knowledge and expertise in the current business processes have been portrayed in the extant literature as capabilities that drive digital transformation in firms (Legner et al. 2017). Segars and Grover (1999) also gestate that the configuration of the IT and organizational resources could lead to superior outcomes when equipped with the capability of strategic planning, while Park and Mithas (2020) posit that firms need to assess their existing capabilities and reconfigure them to a configuration that can eventuate a superior performance. Strategic expertise was addressed in terms of clear and focused vision, rapid scaling, ambidextrous attitude, welldefined strategic goals, IT and business alignment, continuous monitoring and so forth (e. g., (Correani et al. 2020). Digital agility is also required to configure the digital assets with other resources in digital transformation, where continuous sensing and seizing of market opportunities take place to generate value for customers (Teece 2014) and in fact, it is crucial as a company moves through the phases of digital transformation, namely digitization, digitalization and digital transformation (Verhoef et al. 2021). Digital responsibility features leadership, trust and an arrangement of values has been recognized in several studies (Brock and Von Wangenheim 2019; Lokuge et al. 2020). Scholars address digital leadership particularly in reference to the role played by the chief digital officer or chief information officer in a company (Legner et al. 2017). Some commonly addressed values in literature were trust (Li et al. 2018), honesty (Alalwan et al. 2021), transparency (Bharadwaj et al. 2013), resilience (Iivari et al. 2020) and consistency (Correani et al. 2020). Quinn (2016) depicts the widening digital skills gap in the backdrop of digital transformation.

### 3.3.3 The considerations of resource configuration for digital transformation

*"How many configurations do we need to describe all organizational structures?... With our nine parameters, that number would grow rather large... But there is order in the world... a sense of union or harmony that grows out of the natural clustering of elements, whether they are stars, ants or the characteristics of organizations." (Morgan 1979, p. 300)*

The analysis revealed six sub themes that resonate the considerations that are taken in to account when resources are being configured to facilitate digital transformation and they include economic, social, technological, environmental, legal and internal considerations. Brock and Von Wangenheim (2019) portray the lack of agility, resistance to changes from the internal environment of the firm, security risks, unstable technology, insufficient funding, absence of appropriate technology partners and the integration of novel digital technology with the current technology as significant challenges confronted by the firms as they expedite digital transformation. Scholars have addressed concerns over the environmental considerations that must be addressed when implementing digital transformation. While acknowledging the detrimental impacts of digital transformation on the environment, Lokuge et al. (2020) portray the need to arrive at a common space that facilitates the co-existence of both environmental sustainability and digital transformation. Furthermore, the configuration of resources and capabilities to expedite digital transformation also needs to consider the privacy and security of all stakeholders. Papagiannidis et al. (2020) showcase that these concerns have been the top issues for the IT teams, and they gestate that a pressurized firm could be a significant pick out for cyber-attacks such as hacking phishing or malware. Firms must sense the changes in the external environment and adapt to them accordingly (Vial 2019). Bonnet and Westerman (2021) also demonstrate the need for organizations to adapt to the changes eventuated by the global pandemic and reinvent the business while improving their products and services. Laurenza et al. (2018) portray legal and privacy issues as substantial barriers to expediting digital transformation in the health care sector.

### 3.3.4 Contexts of configuration of resources for digital transformation

*"Digital transformation is inevitable for both large-sized enterprises (LSEs) and small and medium-sized enterprises (SMEs)" (Thrassou et al. 2020, p. 152).*

As listed in Table 2, this analysis revealed two major sub themes of contexts namely relative and definite contexts. In this study, the authors identified some salient contextual differences concerning digital transformation literature, specifically in the country and the firm size. Most of the studies reviewed were conducted in developed countries (Laurenza et al. 2018; Quinn et al. 2016). The publications on SMEs and large-scale enterprises (LSEs), particularly in developing countries, were rare. Hence, digital transformation based in developing countries requires further attention. Furthermore, in relation to the size of the firm, digital transformation research based on SMEs are comparatively lower than those based on the LSEs. Scholars also portray some relative aspects in relation to digital transformation, such as ambiguity and complexity (Quinn et al. 2016). As digital transformation eventuates in the introduction of chief digital officers, new reporting lines and many changes across firms, it heightens the ambiguity and complexity of digital transformation (Lanzolla et al. 2020). In terms of consumer behavioural changes, digital technologies tend to influence their decisions, and when the value networks expand, the firms confront more complexity and uncertainty (Vial 2019). Solberg et al. (2020) posit that digital transformation entails several actors who share associations, resulting in a non-realistic type of transformation fuelled with uncertainty, which leads to ambiguity with less clarity within the organization (Quinn et al. 2016).

### 3.3.5 Outcomes of resource configuration for digital transformation

*"Digital transformation is concerned with the changes digital technologies can bring about in a company's business model, which result in changed products or organizational structures or in the automation of processes" (Hess et al. 2016, p. 124).*

This section outlines the effects that business organizations confront from digital transformation and the authors categorize these outcomes into positive and negative. The burgeoning digital technologies that entail the fusion and connectivity of vast information, communication and technologies have posed innumerable impacts on the current business organizations that are required to adapt to the change (Bharadwaj et al. 2013; Verhoef et al. 2021). Competitive advantage (Park and Mithas 2020), new revenue stream generation (Lanzolla et al. 2020), sustainability (Lokuge et al. 2020), profitability (Bonnet and Westerman 2021) and value creation (Laurenza et al. 2018) resonate the sub-theme of efficacy in digital transformation literature. Cenamor et al. (2019) declare how innovation is spurred by digital transformation, while Burden et al. (2018) assert that it stifles innovation in an organization. Authors reflected attributes of enriched data and information where data and information security, access to live usage data, data optimization, new knowledge, knowledge transfer and enhanced information sharing are cited (Euchner 2019). Scholars have also addressed the notion of business process excellence in terms of enhanced decision making, stability, brand enhancement, accuracy, simplified processes, uniformity, improved planning, increased online presence, increased reliability and increased responsiveness and easy detection of obsolete tasks (Bharadwaj et al. 2013; Laurenza et al. 2018). The positive impacts on the environment have been discussed by scholars such as Lokuge et al. (2020), while the detrimental effects of digital transformation on the environment have been addressed by scholars such as Broekhuizen et al. (2021). Literature also address hindrances to business process reconfiguration, erosion of competitiveness, negative word of mouth and increased complexity and exposure of the firm to risks (Cenamor et al. 2019; Lanzolla et al. 2020). Podlesny and Solder (2020) state that contemporary organizations are under substantial pressure to ensure data protection and its responsible use. In fact, as digital technologies subject organizations to rapid changes, ethical issues have become common and many scholars have identified their impact (Broekhuizen et al. 2021; Vial 2019).

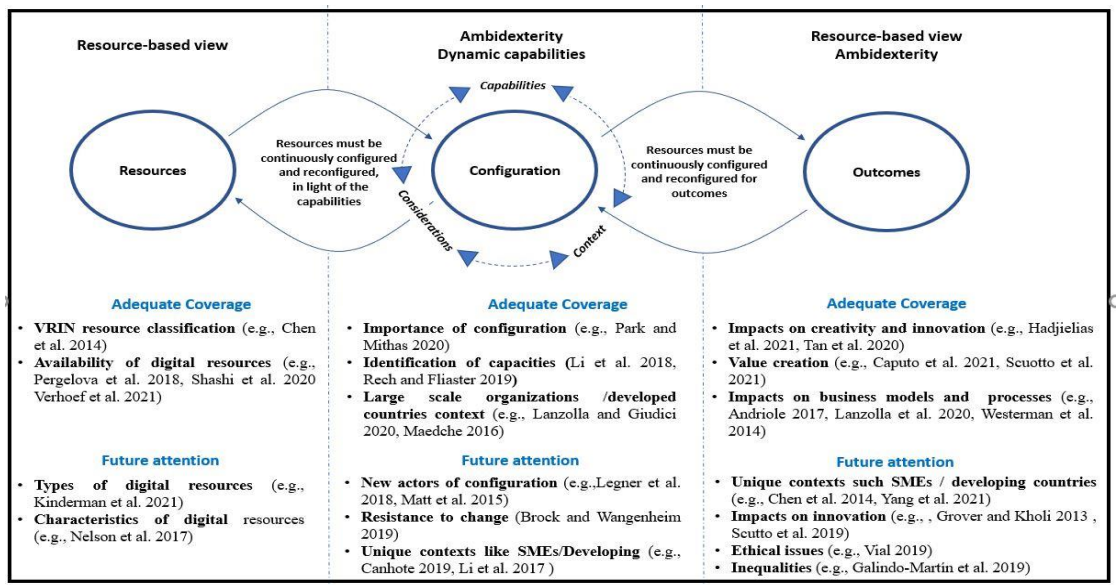

*Figure 3: The RC3O nomological net of digital transformation*

For Table 2. and Figure 3. all the references are not included due to space limitations. Please contact corresponding author to receive a copy of the full reference list.

## 4　Conclusion

This study was conducted to derive a theoretically driven nomological net on digital transformation that illustrates a consolidated view of the multifaceted notion; Digital Transformation. The contributions of this study are threefold. First, it proposes a-priori nomological net that unfolds the current digital transformation research trends as an input-process-output framework and supports the dominant theoretical foundations in contemporary digital transformation literature. Second, the RC3O nomological net proposed by the authors provides the current, and future researchers with a holistic perspective on the patterns and the knowledge gaps present in the discipline. Third, it will enable business organizations to drive digital transformation by considering the critical success factors and taking measures to avoid the inhibitors. The findings of this study portray that there are still less well explored areas in extant literature such as characteristics and types of digital resources, new actors of configuration, environmental considerations impacting configuration, impacts on innovation, inequalities, digital expertise, ethical issues related to digital transformation and unique contexts such as SMEs and / or developing countries. Nonetheless, there are limitations of this study that require consideration. The authors could not address digital transformation of the public sector and used papers exclusively from the IS and management disciplines. The papers published in leading academic outlets were chosen for the study and this might lead to a loss of data published in other scholarly work. The authors encourage prospective researchers to address these voids and add value to current and future research on the notion of digital transformation.